# Effect of Fe substitution on magnetocaloric effect in $La_{0.7}Sr_{0.3}Mn_{1-x}Fe_xO_3$ $(0.05 \le x \le 0.20)$

C. Krishnamoorthi, S. K. Barik, and R. Mahendiran*

Department of Physics and NUS Nanoscience & Nanotechnology Initiative, 2 Science Drive 3, National University of Singapore, Singapore-117542, Singapore.

## Abstract

We have studied the effect of Fe substitution on magnetic and magnetocaloric properties in $La_{0.7}Sr_{0.3}Mn_{1-x}Fe_xO_3$ (x = 0.05, 0.07, 0.10, 0.15, and 0.20) over a wide temperature range (T = 10–400 K). It is shown that substitution by Fe gradually decreases the ferromagnetic Curie temperature ($T_C$) and saturation magnetization up to x = 0.15 but a dramatic change occurs for x = 0.2. The x = 0.2 sample can be considered as a phase separated compound in which both short-range ordered ferromagnetic and antiferromagnetic phases coexist. The magnetic entropy change ($-\Delta S_m$) was estimated from isothermal magnetization curves and it decreases with increase of Fe content from 4.4 $Jkg^{-1}K^{-1}$ at 343 K (x = 0.05) to 1.3 $Jkg^{-1}K^{-1}$ at 105 K (x = 0.2), under ΔH = 5 T. The $La_{0.7}Sr_{0.3}Mn_{0.93}Fe_{0.07}O_3$ sample shows negligible hysteresis loss, operating temperature over 60 K range around room temperature with refrigerant capacity of 225 $JKg^{-1}$, and magnetic entropy of 4 $Jkg^{-1}K^{-1}$ which will be an interesting compound for application in room temperature refrigeration.

* E-mail: phyrm@nus.edu.sg

## 1. Introduction

In recent years, magnetic refrigeration based on magnetocaloric effect has received more attention because it is considered to be more energy efficient and environmentally friendlier compared to the conventional refrigeration based on liquid-vapor compression [1, 2]. The resurgence of interest in magnetocaloric materials is partly due to the discovery of a giant magnetic entropy change in $Gd_5Si_2Ge_2$ alloys in late 1990's [3]. The magnetocaloric effect relies on the fact that spin entropy of a magnetic material decreases upon application of an external magnetic field and this reduction in magnetic entropy is compensated by an increase in the lattice entropy resulting in increase of temperature of the sample. Conversely, when the magnetic field is removed adiabatically, magnetic spins tend to randomize which leads to increase in the magnetic entropy and decrease in the lattice entropy and hence lowering temperature of the sample. It was shown that the temperature of the $Gd_5Si_2Ge_2$ decreases as much as 15 K compared to 11 K with Gd metal under $\Delta H = 5$ T [3]. This giant temperature change in $Gd_5Si_2Ge_2$ compound is due to magnetic-field induced first-order meta magnetic transition, in which the paramagnetic (PM) phase transforms into ferromagnetic (FM) phase abruptly above a critical value of magnetic field at constant temperature. Morelleon *et al.* [4] showed that the metamagnetic transition in this compound is accompanied by structural transition under magnetic field.

Following the discovery of giant magnetocaloric effect (MCE) in Gd-Si-Ge alloys, a great effort has been done theoretically and experimentally to design a new refrigerator and search for an effective giant magnetic refrigerants for expanding the working temperature near room temperature. Subsequently, a large magnetic entropy change was reported in other alloys such as $MnFeP_{0.45}As_{0.55}$ [5], $Fe_{49}Rh_{51}$ [6], $La(Fe, Si)_{13}$ [7] and $Mn_{1-x}Fe_xAs$ [8]. Although these are some of the good magnetic refrigerants to obtain magnetic refrigeration at room temperature, they are limited in use by the cost and environmental issues of the ingredient elements, such as As, and engineering aspects. Alternative to these alloys, perovskite manganites of the general formula $R_{1-x}A_xMnO_3$ (R-rare-earth, A-alkali-earth) are also being explored for possible high magnetic entropy change ($\Delta S_m$) values [9, 10]. Manganites attract particular attention due to their higher chemical stability, high resistivity resulting in lower eddy current heating, lower cost, simple preparation methods and more importantly the ability to tailor their magnetic transition temperatures close to room temperature by R-site or Mn-site substitution. These properties are much deserved, in addition to high MCE, to realize affordable room temperature magnetic refrigeration for domestic applications.

$La_{0.7}Sr_{0.3}MnO_3$ is one of the extensively studied manganites which undergoes a paramagnetic metal to a ferromagnetic metal transition around $T_C$ = 375 K and it shows a $\Delta S_M$ of 4.6 $Jkg^{-1}K^{-1}$ under $\mu_0H$ = 5 T around its $T_C$ [11]. The ferromagnetic transition of this compound can be brought down to room temperature either by partial replacement of the bigger ionic size $La^{3+}$ by smaller size $Pr^{3+}$ or $Nd^{3+}$ ion or by partial replacement of Mn ions by other transition metal ions such as M = Ni, Cr, Fe, Zn etc [12]. It is observed the Curie temperature and the magnetic entropy of $La_{0.7}Sr_{0.3}MnO_3$ decreases by small amount of $Al^{3+}$ (5 %) substitution at Mn site whereas the magnetic entropy is not severely affected by similar quantity of $Ti^{4+}$ substitution [11]. In this context, Fe substitution in $La_{0.7}Sr_{0.3}MnO_3$ deserves a particular attention due to magnetic nature of the Fe ion compared to non-magnetic Al and Ti. In one of the earliest work on the Fe substituted manganites, Leung *et al.* [13] studied magnetic properties of $La_{1-x}Pb_xMn_{1-y}Fe_yO_3$ and concluded that antiferromagnetic (AFM) $Fe^{3+}$-O-$Fe^{3+}$ and $Fe^{3+}$-O-$Mn^{3+}$ interactions competed with the ferromagnetic $Mn^{3+}$-O-$Mn^{4+}$ double exchange interactions. Later,

Ahn *et al.*[14] found that Fe doping destroys FM metallic state in $La_{0.7}Ca_{0.3}Mn_{1-x}Fe_xO_3$. The Mössbauer studies on the $La_{0.75}Ca_{0.25}Mn_{0.98}Fe_{0.02}O_3$ indicate coexistence of PM and FM phases well below $T_C$, and the fraction of the PM phase decreases with lowering the temperature and becomes zero at low temperatures [15]. Damay *et al.*[16] found several orders of decrease in the resistivity under magnetic field in $Sm_{0.7}Sr_{0.3}Mn_{1-x}Fe_xO_3$ (x = 0.03) compared to x = 0.0. Although $La_{0.7}Sr_{0.3}Mn_{1-x}Fe_xO_3$ series ($0 \leq x \leq 0.1$) had been studied by many groups in the past from the perspective of dc magnetoresistance and magnetization [17, 18], a detail study on MCE over a wide composition is currently lacking and this is the motivation of the present work. During the course of our work, MCE study on one of the compositions ($La_{0.7}Ca_{0.3}Mn_{0.9}Fe_{0.1}O_3$) was reported [19]. Here, we report magnetic and magnetocaloric (MCE) properties of $La_{0.7}Sr_{0.3}Mn_{1-x}Fe_xO_3$ (x = 0.05, 0.07, 0.1, 0.15 and 0.2) over a wide temperature range under various magnetic field strengths.

**2. Materials and Methods**

The pre-determined stoichiometric compounds of $La_{0.7}Sr_{0.3}Mn_{1-x}Fe_xO_3$ system with x= 0.05, 0.07, 0.10, 0.15 and 0.20, were prepared by standard solid state reaction method from high pure $La_2O_3$ (99.9%), $SrCO_3$ (99.9+%), $Fe_2O_3$ (99.98%) and $Mn_2O_3$ (99.9%) precursor compounds. The stoichiometric precursor compounds were ground together thoroughly and calcined at 900 $^{o}$C for 20 h. In succession, the resultant pristine samples were ground into fine powder and heated at 1050 $^{o}$C for 20 h. The later procedure was repeated for two times with 50 $^{o}$C heat increment for each time and finally the pressed pellets were sintered at 1200 $^{o}$C for 20 h. The temperature and magnetic field dependent magnetization of all the samples were measured by commercial vibrating sample magnetometer (VSM), from Quantum Design Inc.USA, under magnetic fields up to $\mu_0H$= 5 T. The isothermal magnetization data were collected during ascending and descending the magnetic field at 5 K intervals around the magnetic transition. The $\Delta S_m$ was estimated from field ascending magnetic isotherms. The zero field cooled (ZFC) magnetization data was collected by cooling the sample down to 10 K in the absence of magnetic field and then the data were collected while warming the sample under the required magnetic field. For the field cooled (FC) data, the sample was cooled down to 10 K under required field and then data were collected while warming the sample under the same field strength.

Magnetic entropy change ($\Delta S_m$) of the samples was estimated from magnetic isotherms. The $\Delta S_m$ of a material at an average temperature $T_{av}= [T_u+T_l]/2$ under the magnetic field change of $\Delta H= H_f-H_i$, from the two magnetic isotherms at $T_u$ and $T_l$, could be estimated using the following standard Maxwell equation [20]:

$$\Delta S_m(T_{av})_{\Delta H} = \int_{H_i}^{H_f} \left[\frac{\partial M(T,H)}{\partial T}\right]_H dH \quad (1)$$

The above equation can be converted into its equivalent numerical summation, using trapezoidal rule, as shown below

$$\Delta S_m(T_{av})_{\Delta H} = \frac{1}{2\delta T}\left(\delta M_1 \delta H_1 + 2\sum_{k=2}^{n-1} \delta M_k \delta H_k + \delta M_n \delta H_n\right) \quad (2)$$

here $\delta T$ is the difference between the temperature of the upper temperature ($T_u$) magnetic isotherm and the lower temperature ($T_l$) magnetic isotherms, n is the number of points measured for each of the two isotherms with the magnetic field changing from $H_i$ to $H_f$ at field interval $\delta H= \Delta H/(n-1)$, and $\delta M_k= [M(T_u)_k-M(T_l)_k]$ is the difference in the magnetization at $T_u$ and $T_l$ at each magnetic field step (k) from 1 to n. In order to make summation easier we have obtained the data at constant magnetic field and temperature intervals so that $\delta H$ and $\delta T$ become constants.

The above procedure is extended to isotherms at different temperatures to obtain $\Delta S_M$ at required $T_{av}$.

**4. Results**

Figure 1(a) shows the temperature dependent magnetization (M−T) for $La_{0.7}Sr_{0.3}Mn_{1-x}Fe_xO_3$ system (x = 0.05, 0.07, 0.10, 0.15, & 0.20) measured under H = 100 Oe. All the samples undergo PM to FM transition upon lowering the temperature as suggested by the rapid increase of magnetization around the ferromagnetic Curie temperature ($T_C$). All the samples except x = 0.2 show a sharp ferromagnetic transition at $T_C$. The $T_C$, determined as the inflection point in derivative of the M−T curve, decreases with increasing Fe content ($T_C$ = 343, 296, 260, 180, & 94 K for x = 0.05, 0.07, 0.10, 0.15, & 0.20 samples, respectively). The magnetic moment at the lowest temperature (T = 10 K) also decreases with increase of the Fe content.

Figure 1(b) shows the temperature dependent inverse susceptibility ($1/\chi$) for all the compositions along with the Curie-Weiss law fits ($1/\chi = (T-\theta_p)/C$). While the Curie-Weiss law fit the inverse susceptibility data of x = 0.05 rather perfectly, we notice that other compositions show a deviation from the linearity as the temperature approaches the $T_C$. It is to be noted that the deviation from the Curie-Weiss fit in x = 0.2 starts below 250 K *i.e.,* much above its $T_C$ (= 94 K) estimated from the M−T data. The paramagnetic Curie temperature ($\theta_p$) obtained from the Curie-Weiss fit, the $T_C$ determined from the M−T curve and estimated the effective magnetic moment ($P_{eff}$) from the estimated Curie constant (C) and theoretical $P_{eff}$ (for spin only Mn, Fe moment) values are listed in table 1 for easy comparison.

Figure 2 shows the magnetic field dependence isothermal magnetization data, M−H, of the $La_{0.7}Sr_{0.3}Mn_{1-x}Fe_xO_3$ at 10 K measured up to $\mu_0H$ = 5 T. The M−H curves show a rapid increase at low fields and then almost saturation at higher fields above 0.5 T for $x \leq 0.15$ which suggests that the long range ferromagnetic order is preserved at low temperature in these compositions. On the other hand, M–H curve of x = 0.2 shows a rapid increase at low fields followed by a gradual increase without saturation even at higher fields indicates that the magnetic ground state dramatically changes from the long-range ferromagnetism to a short-range ferromagnetism when the Fe content increases from 15 % to 20 %. The saturation magnetization ($M_S$) decreases with increase of Fe content. The $M_S$ was estimated by extrapolating the linear part of inverse magnetic field [$(\mu_0 H)^{-1}$] dependent magnetization curves as shown in the inset of Fig. 2. The estimated $M_S$ along with their theoretical values for all the compounds are listed in table 1. The experimentally obtained $M_S$ values of x = 0.05, 0.07, & 0.10 are slightly lower but close to their theoretical values. However, it is much smaller than the theoretical value for x = 0.2.

The Arrott plots (H/M vs $M^2$ or $1/\chi$ vs $M^2$) are generally used to determine order of the magnetic phase transitions [21, 22]. In a simple paramagnetic to ferromagnetic transition, the linear fit to the high $M^2$ data of the Arrott plot intercepts on the ordinate ($1/\chi$) at zero for $T = T_C$ since by definition $1/\chi = 0$ at $T_C$. The existence of spontaneous magnetization is inferred from a rapid increase of $M^2$ at near the origin of the ordinate, i.e., at very low H/M value. Further, samples which exhibit second-order phase transition show a positive slope at all points of $M^2$ in the Arrott plot, at and below $T_C$, whereas samples which exhibit first-order phase transition show a negative slope for certain range of $M^2$ values [23]. The coefficient of $M^3$ term in the magnetic equation of state $H = AM + BM^3$ is expected to be negative for first order phase transition. We have shown the Arrott plot of $La_{0.7}Sr_{0.3}Mn_{1-x}Fe_xO_3$ (x= 0.07, 0.10, and 0.20) in figure 3 at 5 K intervals around their respective $T_C$. Arrott plots of other samples are similar to x = 0.10 sample. The Arrott plots of all the samples show positive slope (B) at all the temperature and magnetic fields and thus suggest a second order para- to ferromagnetic transition. The $T_C$ estimated from the M−T curves and the Arrott plots are in close to each other, e.g., the M−T curve of x = 0.07 sample yields $T_C$ = 296 K and the Arrott plots also give almost the same value of $T_C$, between 295 to 300 K. While compositions from x = 0.05 to x = 0.15 exhibit a similar behavior, the x = 0.20 sample shows a small spontaneous magnetization ($M^2$ intercept). The Arrott plot of x = 0.20 sample resembles that of cluster-glass samples which have short range ferromagnetic interactions as seen in $(Fe_{0.7}Mn_{0.3})_{0.75}P_{15}C_{10}$ amorphous alloys [24]. The cluster-glass nature is also seen in x = 0.20 sample through the irreversibility in ZFC and FC magnetization just below $T_C$ under H = 100 Oe as shown in figure 4. Below the irreversibility temperature ($T_{ir}$), the ZFC and FC magnetization bifurcate and the ZFC curve shows a broad cusp whereas the FC curve nearly saturates. The $T_{ir}$ shifts down to 50 K under $\mu_0H$ = 1T but does not vanish. These behavior suggest that the low temperature phase is not a homogenous ferromagnet. The low temperature phase can be considered as a mixture of FM and AFM domains and this point will be discussed later.

Figure 5 shows the isothermal magnetization curves for (a) x = 0.05, (b) x = 0.15 and (c) x = 0.2 samples around their $T_C$. The x = 0.05 samples shows typical M–H behavior of the ferromagnet, with a rapid increase at low fields and nearly saturation at high fields. However, the approach to the saturation becomes rather slow in x = 0.15 and no saturation in x = 0.2. The temperature dependent $\Delta S_m$ of all the samples were shown in figures 6 and 7 over a wide temperature range around their respective $T_C$ under H = 1, 3 and 5 T. All the curves show a broad maximum of $\Delta S_m$ around their respective $T_C$. The value of $\Delta S_m$ peak increases with the field and the peak position remain nearly unaffected. The peak value of $\Delta S_m$ (= 4.4 $Jkg^{-1}K^{-1}$) of x = 0.05 under $\mu_0H$ = 5 T is very close to value reported in the parent compound $La_{0.7}Sr_{0.3}MnO_3$ ($\Delta S_m$ = 4.6 Jkg$^{-1}$K$^{-1}$) [11] but is almost double the value of $La_{0.7}Sr_{0.3}MnO_3$ reported by others [25, 26]. As the Fe content increases the magnitude of $\Delta S_m$ decreases under a given field strength. The peak value of $\Delta S_m$ under $\mu_0$H= 5 T field is 4.4, 4.0, 3.1, 2.5, and 1.3 Jkg$^{-1}$K$^{-1}$ for x = 0.05, 0.07, 0.10, 0.15 and 0.20 samples at 343, 297, 259, 182, 127 and 107 K respectively. The value of $\Delta S_m$ for $\mu_0$H= 2 and 5 T are listed in table 2 along with related compounds [27] for easy comparison. It is to be noted that $\Delta S_m$ = −2.22 J Jkg$^{-1}$K$^{-1}$ in $La_{0.7}Sr_{0.3}Mn_{0.9}Fe_{0.1}O_3$ under 3 T at $T_C$ = 250 K and is almost double the value of $\Delta S_m$ (= −1.18 Jkg$^{-1}$K$^{-1}$) at $T_C$ = 113 K reported for the related compound $La_{0.67}Ca_{0.33}Mn_{0.9}Fe_{0.1}O_3$ at the same field strength [19].

The relative cooling power (RCP) or refrigerant capacity of a material gives an estimate of quantity of the heat transfer between the hot and cold sinks during one refrigeration cycle and is the area under the $-\Delta S_m$ versus T curve between the temperatures at the full-width at half-maximum (FWHM) of the curve, as shown in Fig.8. The RCP values are estimated by its numerical equivalent: product of the peak value of $-\Delta S_m$ and FWHM ($\delta T_{FWHM}$) of the $-\Delta S_m$ versus T curve (RCP = $-\Delta S_m * \delta T_{FWHM}$). We have measured the peak value from the base line of the curve. The RCP values of all the compounds under 2 T and 5 T fields are listed in table 3 along with related compounds. The obtained RCP values are little higher than the parent compound [11]. The RCP value increases linearly with an applied field for all the present compounds as shown in the inset of Fig.9 and they also increases linearly with Fe content up to x= 0.15 and then decreases with further increase of Fe as shown in the Fig.9.

The operative temperature range of a magnetic refrigerant is an important parameter and gives the temperature range in which the refrigerant could be used efficiently and it is the temperature region between the $T_{hot}$ and $T_{cold}$ temperatures as shown in Fig.8. The operative temperature (OT) range ($T_{hot}$-$T_{cold}$) of the $La_{0.7}Sr_{0.3}Mn_{1-x}Fe_xO_3$ system increases with field as well as Fe content as shown in fig. 9. The OT range of all the compounds under H = 2 T as well as 5 T field are listed in table 3. The OT of x = 0.07 is 41 K under H = 2 T which is higher than the parent compound (30 K) under the same field [11]. Among all the samples, the x = 0.2 sample shows highest OT of 98 and 130 K under 2 T and 5 T, respectively.

Figure 10 shows field dependence magnetization of $La_{0.7}Sr_{0.3}Mn_{0.93}Fe_{0.07}O_3$ sample at different temperatures, around its $T_C$, both in ascending as well as descending field. All the curves show no hysteresis in the magnetization and similar curves were obtained for other samples. The inset shows enlarged view of the magnetic isotherm, at 300 K, between 1.4 and 1. 8 T field and it clearly shows negligible field dependence magnetization hysteresis. The temperature dependence magnetization of x = 0.07 sample around its $T_C$ were shown in Fig. 11 for both warming as well as cooling cycles. The sample exhibits negligible thermal hysteresis around its $T_C$. A similar behavior is observed for other samples.

## 5. Discussion

The gradual decrease of $T_C$ as well as the saturation magnetization ($M_s$) with increase of Fe content in $La_{0.7}Sr_{0.3}MnO_3$ indicates weakening of the double exchange ferromagnetic interactions associated with the itinerancy of $e_g^{\,1}(S=1/2)$ holes in the background of localized $t_{2g}^{\,3}(S=3/2)$ electrons. The substitution of $Fe^{3+}(3d^5{:}t_{2g}^{\,3}e_g^{\,2})$ ion for $Mn^{3+}$ $(3d^4{:}t_{2g}^{\,3}e_g^{\,1})$ ion disrupts charge transfer along the $Mn^{3+}$-O-$Mn^{4+}$ net work as well as reduces the density of itinerant $e_g$ electrons. Because the $e_g$ and $t_{2g}$ electrons on $Fe^{3+}$ are localized on the Fe ion and such electrons participate in superexchange antiferromagnetic interactions with their neighboring Mn and Fe ions in the lattice [13]. The antiferromagnetic interactions compete with ferromagnetic interactions in the Fe-substituted compounds and this lead to decrease of $M_s$ as well as $T_C$. Nevertheless, our magnetization data suggest long range ferromagnetic order is maintained up to x = 0.15 and it dramatically changes to short-range ferromagnetic order for x = 0.20. It is important to refer to other microscopic studies. Neutron depolarization study by Yusuf *et al.,* [28] indicated preservation of ferromagnetic domains up to 10 % Fe substitution in $La_{0.67}Ca_{0.33}MnO_3$ but they had not investigated for higher Fe content. Barandiarán *et al.,*[29] studied $La_{0.7}Pb_{0.3}Mn_{1-x}Fe_xO_3$ ($0 \leq x \leq 0.3$) using different experimental techniques (magnetization, Mössbauer spectroscopy, wide-angle polarized, depolarized neutron scattering and small-angle neutron scattering) and concluded that short-range ordered ferromagnetic (FM) and antiferromagnetic (AFM) clusters of different sizes coexist in x = 0.2 sample. The competition among these short-range FM and AFM phases result in a cluster-glass like behavior for $x \geq 0.2$ compositions. Thus, the sudden decrease in the spontaneous magnetization of x = 0.20 in our compounds is certainly connected to the breakdown of the long range ferromagnetic order and transition into cluster glass like state in which both FM and AFM domains coexist. In x = 0.20, the observed irreversibility in ZFC and FC magnetization curves just below $T_C$ (Fig.4) and small spontaneous magnetization at 10 K (inset of Fig.2) and behavior of Arrott plots (Fig.3) and our electrical resistivity measurements support this cluster glass scenario. This sample shows insulating behavior down to 10 K whereas the x = 0.15 showed a clear insulator-metal transition at around 80 K (not shown here).

As the magnetic inhomogeneity increases with Fe content, the peak value of the magnetic entropy change ($-\Delta S_m$) decreases and the $-\Delta S_m$ spreads over a wide range of temperature around their $T_C$'s. In the cluster-glass compound (x = 0.20) the $\Delta S_m$ spread over 300 K range under 5 T field as shown in Fig.7. The destruction of native long range FM interactions increases with Fe content. This leads to reduction of $M_s$ and a broad temperature dependence magnetic transition and hence dM/dT decreases with increase of Fe content. Hence, the $\Delta S_m$ decreases and spread over a wide temperature range with increase of Fe content. The observed large magnetic entropy change in the $La_{0.7}Sr_{0.3}Mn_{1-x}Fe_xO_3$ system is reversible in both field (Fig.10) as well as thermal (Fig.11) cycles. Further, the $\Delta S_m$ exist around room temperature in x = 0.05 to 0.10 compounds. Particularly, the x = 0.07 sample shows large $-\Delta S_m$ at 297 K. Reversible large $\Delta S_m$ and large $\Delta S_m$ at room temperature are the two properties most sought in refrigerant to design a reliable, efficient room temperature magnetic refrigerator [30]. The materials which exhibit magnetic field or thermal cycle hysteresis in $\Delta S_m$ show less refrigeration efficiency. The negligible thermal and magnetic field hysteresis in magnetization for $La_{0.7}Sr_{0.3}Mn_{1-x}Fe_xO_3$ system over a wide temperature range around room temperature makes this system very useful as magnetic refrigerant for room temperature magnetic refrigerator. In addition, relatively cheaper cost and easy preparation methods of these materials compared with the classic room temperature magnetic refrigerant, rare-earth Gd metal, is an additional advantage. Furthermore, the resistivity of the present compounds, both in the absence and presence of magnetic field, are found to be very high (> $10^2$ Ω-cm) over the operating temperature range. The higher resistivity of the samples reduces the eddy current losses in magnetic field cycling. Though the $\Delta S_m$ values of x = 0.07 sample is inferior to the Gd metal, the other factors such as cost, $\Delta S_m$ reversibility and high resistivity are superior to Gd.

In lanthanum based manganites, the observed temperature dependence magnetic entropy change have been fitted with models based on Weiss molecular mean field theory [31] or Landau theory of phase transition [[32]32]. Among both the models, the model based on Landau theory fit the data very closely in lanthanum based manganites. In order to understand the observed temperature dependence magnetic entropy change of the samples, of particular interest in x =

0.07 sample, we have tried to fit the data to the model based on Landau theory with Gibb's free energy

$$G(T,M) = G_0 + \frac{1}{2}AM^2 + \frac{1}{4}BM^4 - MH \quad (3)$$

where A and B are temperature dependence coefficients. The coefficient B includes elastic and magnetoelastic terms. By differentiation of the magnetic part of the above equation with respect to temperature, the magnetic entropy change is obtained:

$$S_m(T,H) = -\frac{1}{2}\frac{\partial A}{\partial T}M^2 - \frac{1}{2}\frac{\partial B}{\partial T}M^4 \quad (4)$$

By assuming equilibrium condition of Gibb's free energy: $\frac{\partial G}{\partial M} = 0$, the linear fit to high $M^2$ data of Arrott plots yield the A and B parameters. The temperature dependence A and B coefficients of x = 0.07 samples are shown in inset of Fig. 12. The A coefficient varies from positive to negative through zero at $T_C$. The B coefficient is positive and it goes through a minimum at just above the $T_C$. For a simple ferromagnet, B is a positive constant and $-\Delta S_m$ is expected to show a narrow peak at $T_C$. It has been suggested that variation of B with temperature implies a substantial temperature dependent contributions from magnetoelastic coupling which we see in our sample (x = 0.07). The temperature dependence $\Delta S_m$ is calculated using temperature dependence A and B coefficients and magnetization data using eq.(4) and is shown for x = 0.07 sample (solid line) along with experimental data under H= 5 T (solid circles). Both experimental and theoretical curves are close to each other which suggest that the magnetic entropy change in this compound can be understood through the Landau theory of phase transition.

## 6. Summary and Conclusion

The magnetic and magnetocaloric properties $La_{0.7}Sr_{0.3}Mn_{1-x}Fe_xO_3$ (x= 0.05, 0.07, 0.10, 0.15, & 0.20) system have been studied systematically over a wide temperature range. The Fe substitution decreases the $T_C$ and the saturation magnetization. Our magnetization results indicate coexistence of ferromagnetic and antiferromagnetic clusters in x = 0.2. The temperature dependence magnetic entropy change ($\Delta S_m$) of all the samples showed a maximum around their respective $T_C$ and its magnitude decreases from 4.4 $Jkg^{-1}K^{-1}$ (x = 0.05) to 1.3 $Jkg^{-1}K^{-1}$ (x = 0.2) with increase of Fe content, under 5 T field. However, the operating temperature range and refrigerant capacity of the present magnetic refrigerants increases with Fe doping. The $La_{0.7}Sr_{0.3}Mn_{0.93}Fe_{0.07}O_3$ sample shows operating temperature over 60 K range around room temperature with refrigerant capacity of 225 J/ kg and magnetic entropy of 4 $Jkg^{-1}K^{-1}$ which will be an interesting compound for application in room temperature refrigeration.

**Acknowledgments**

RM acknowledges the Deputy President (Office of Research and Technology) office, NUS for supporting this work through the grant no. R-144-000-197-123.

## Figure Captions:

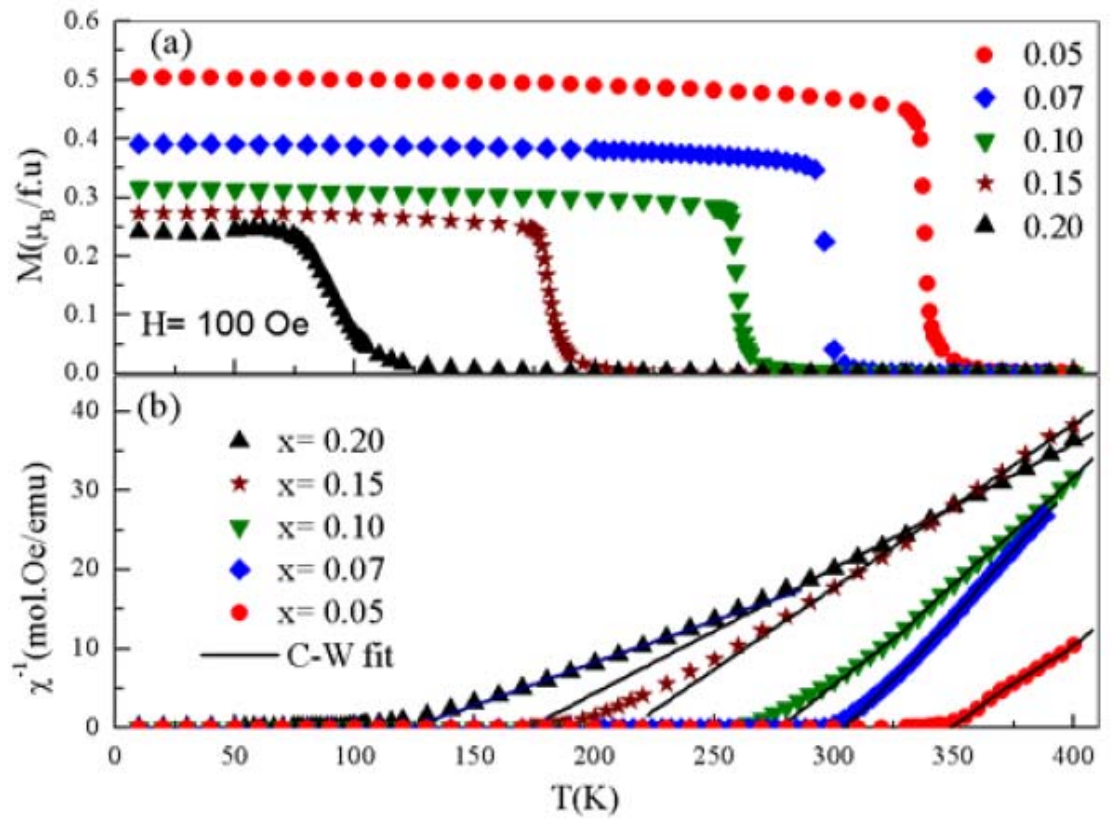

**Fig.1**: The temperature dependent (a) magnetization and (b) inverse susceptibility of the $La_{0.7}Sr_{0.3}Mn_{1-x}Fe_xO_3$ (x= 0.05, 0.07, 0.10, & 0.20) system under an applied field of 100 Oe. The solid lines in (b) are Curie-Weiss (C–W) law fit to the data in paramagnetic regime. Note that C–W law is fitted for two linear regions, one from 400 K to 270 K and another one from 250 K to 150 K in x= 0.20.

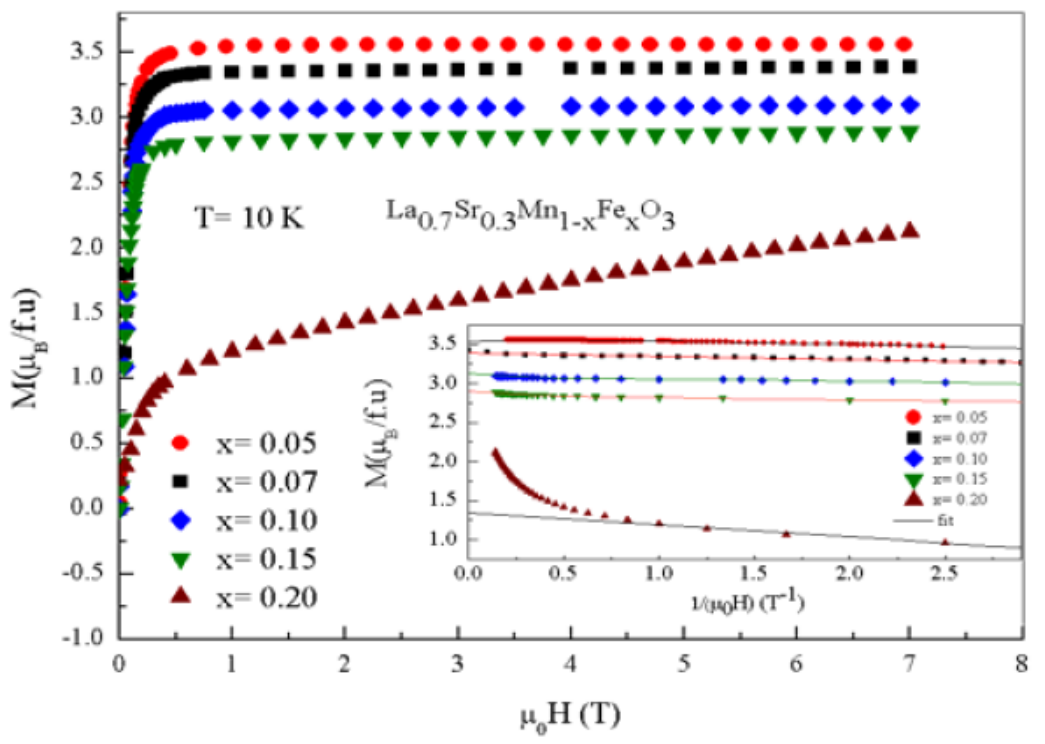

**Fig. 2**: The magnetic field dependent magnetization for the x = 0.05, 0.07, 0.10, 0.15, and 0.20 samples at 10 K. The inset shows their respective inverse magnetic field $[(\mu_0H)^{-1}]$ dependent magnetization and linear fit to the high $(\mu_0H)^{-1}$data (solid line).

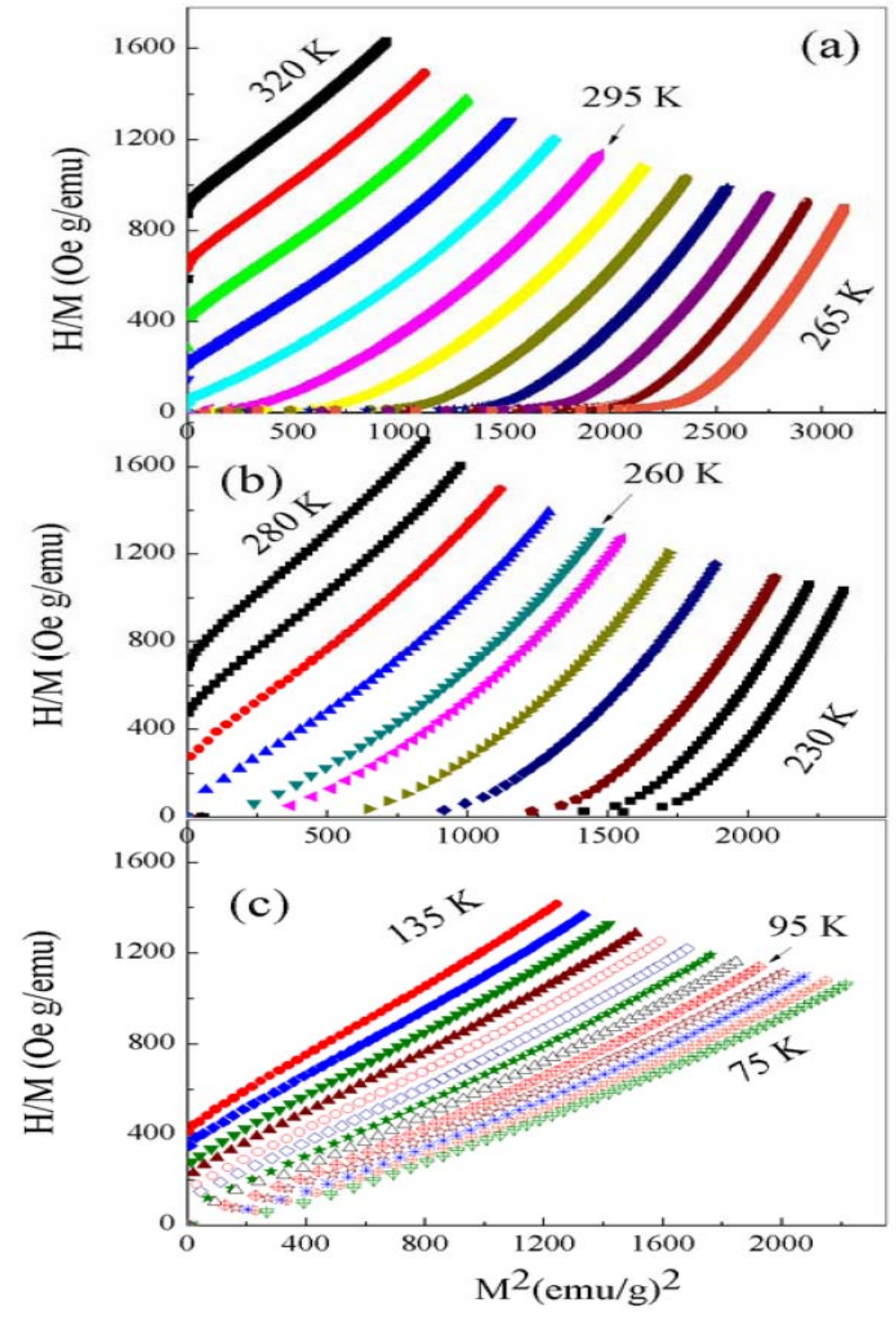

**Fig. 3**: The Arrott plots (H/M vs. $M^2$) of the $La_{0.7}Sr_{0.3}Mn_{1-x}Fe_xO_3$ samples around their respective $T_C$ at 5 K intervals. Figure (a), (b) and (c) represents x = 0.07, 0.10, & 0.20 respectively.

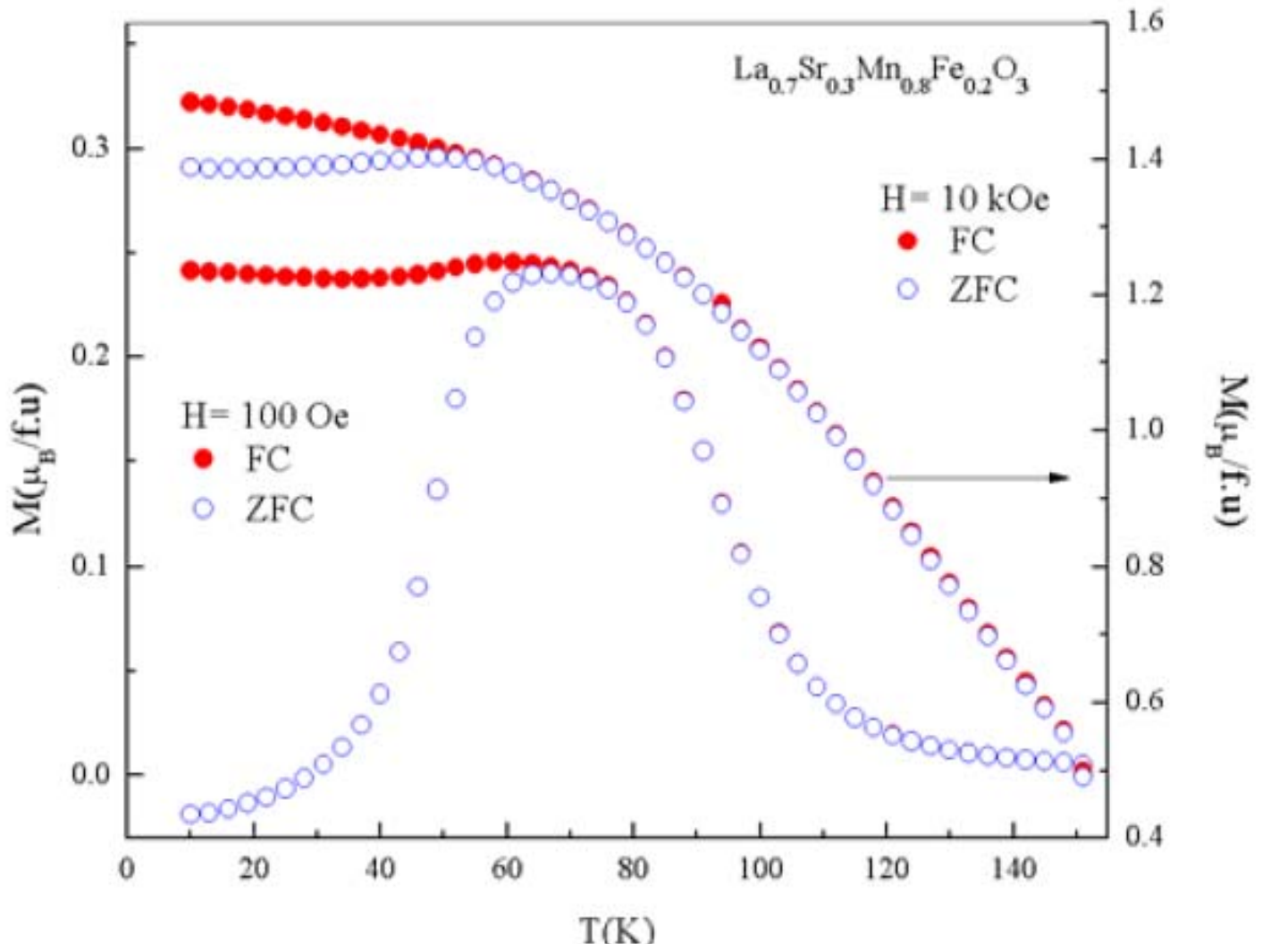

**Fig. 4**: The temperature dependence magnetization of the $La_{0.7}Sr_{0.3}Mn_{0.80}Fe_{0.20}O_3$ sample in ZFC and FC condition under 100 Oe (left ordinate) and 10 kOe field (right ordinate).

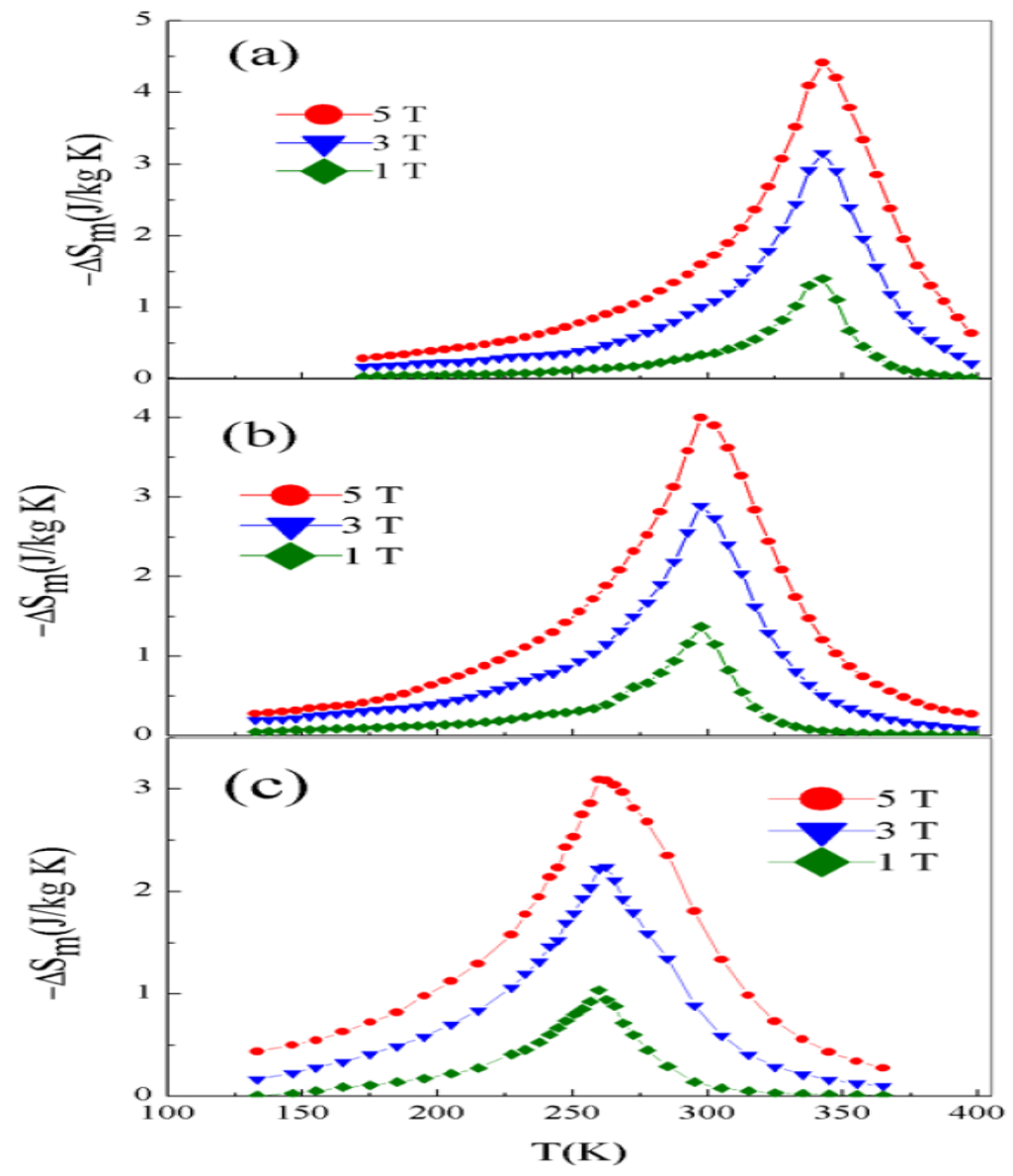

**Fig.5**: The isothermal magnetization (M−H) curves of $La_{0.7}Sr_{0.3}Mn_{1-x}Fe_xO_3$ samples around their Curie temperature ($T_C$) at 5 K intervals. Figure (a), (b) and (c) represents x= 0.05, 0.15 and 0.20, respectively.

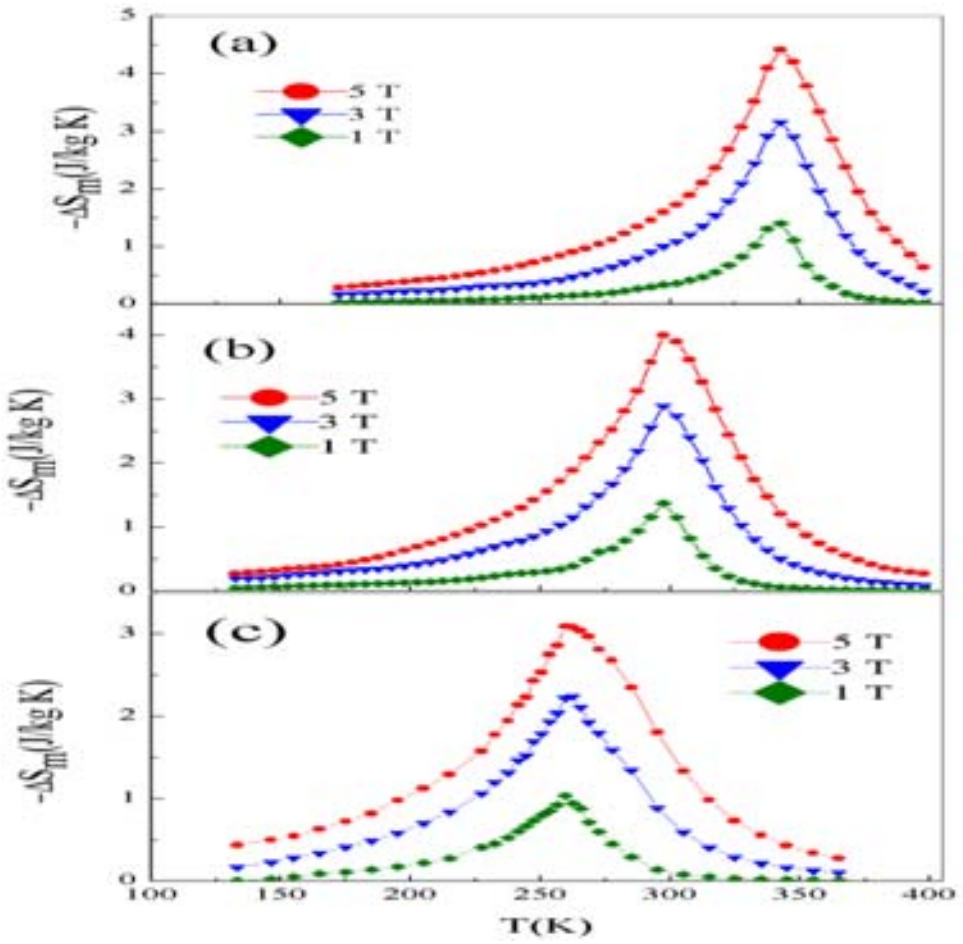

**Fig. 6**: The temperature dependent magnetic entropy change ($-\Delta S_M$) of $La_{0.7}Sr_{0.3}Mn_{1-x}Fe_xO_3$ samples around their respective magnetic phase transition for (a) x = 0.05, (b) 0.07, and (c) 0.10. The lines are guide to eye. Note that the peak temperature is unaltered with increase of the magnetic field.

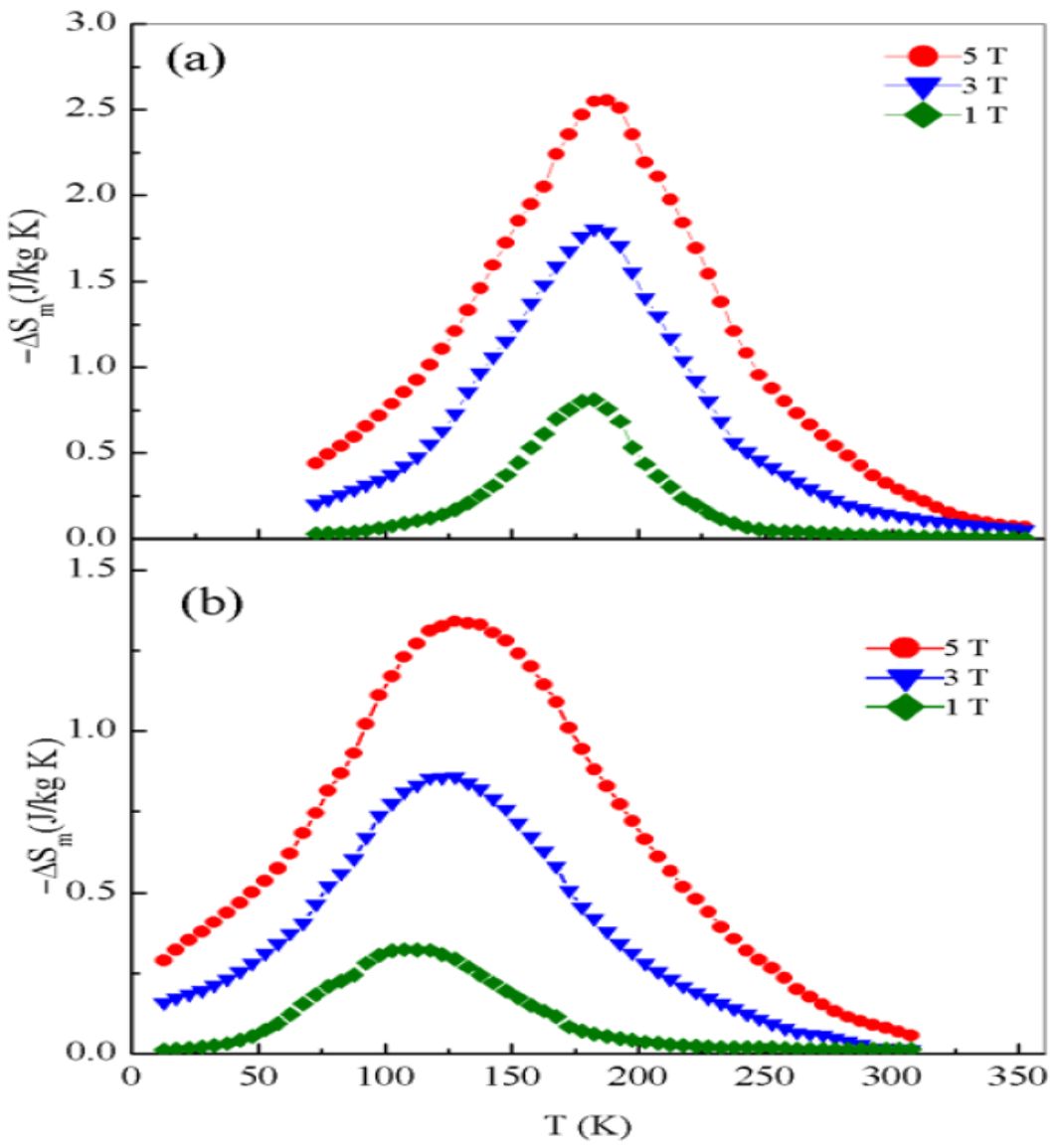

**Fig. 7**: The temperature dependent magnetic entropy change ($-\Delta S_M$) of $La_{0.7}Sr_{0.3}Mn_{1-x}Fe_xO_3$ around their magnetic phase transition for (a) x = 0.15 and (b) x = 0.20. The lines are guide to eye.

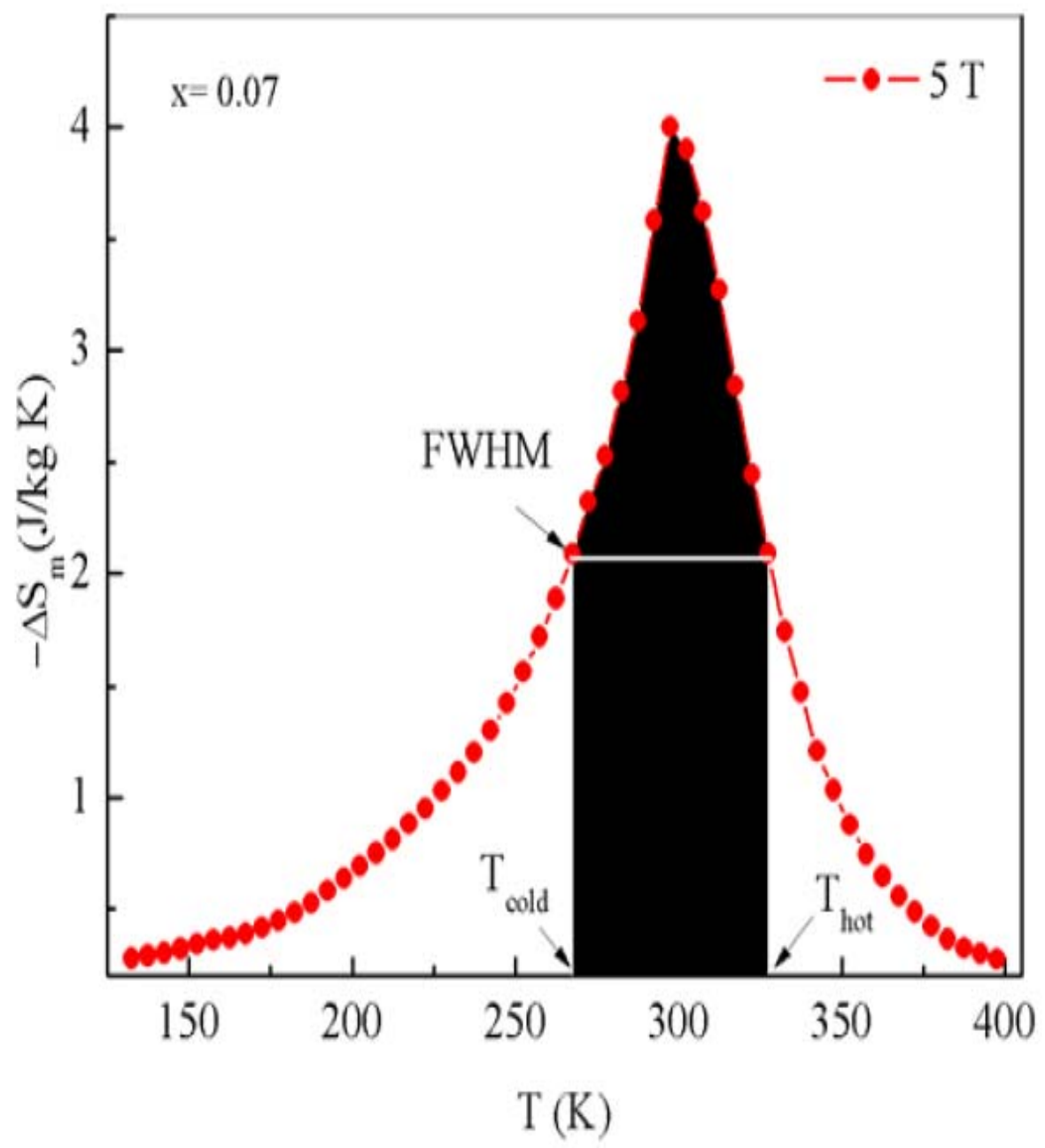

**Fig.8**: The temperature dependent $-\Delta S_m$ of $La_{0.7}Sr_{0.3}Mn_{0.93}Fe_{0.07}O_3$ under 5 T field. The shaded area is the relative cooling power. The white line shows full-width at half maximum.

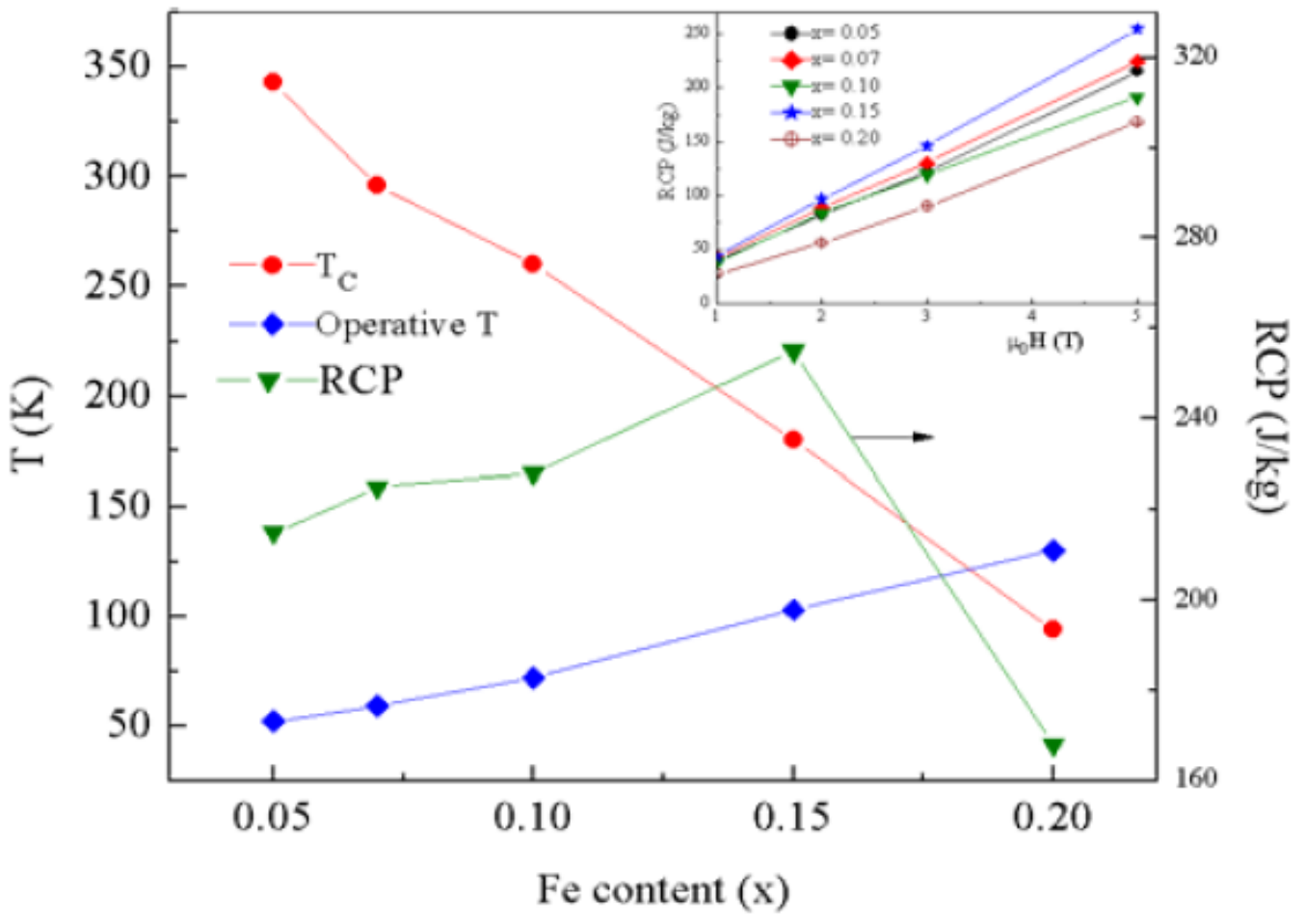

**Fig.9**: Fe content (x) dependent $T_C$, operative temperature range and relative cooling power (right ordinate) of the $La_{0.7}Sr_{0.3}Mn_{1-x}Fe_xO_3$ system. The inset shows magnetic field dependence RCP for all the samples under 5 T.

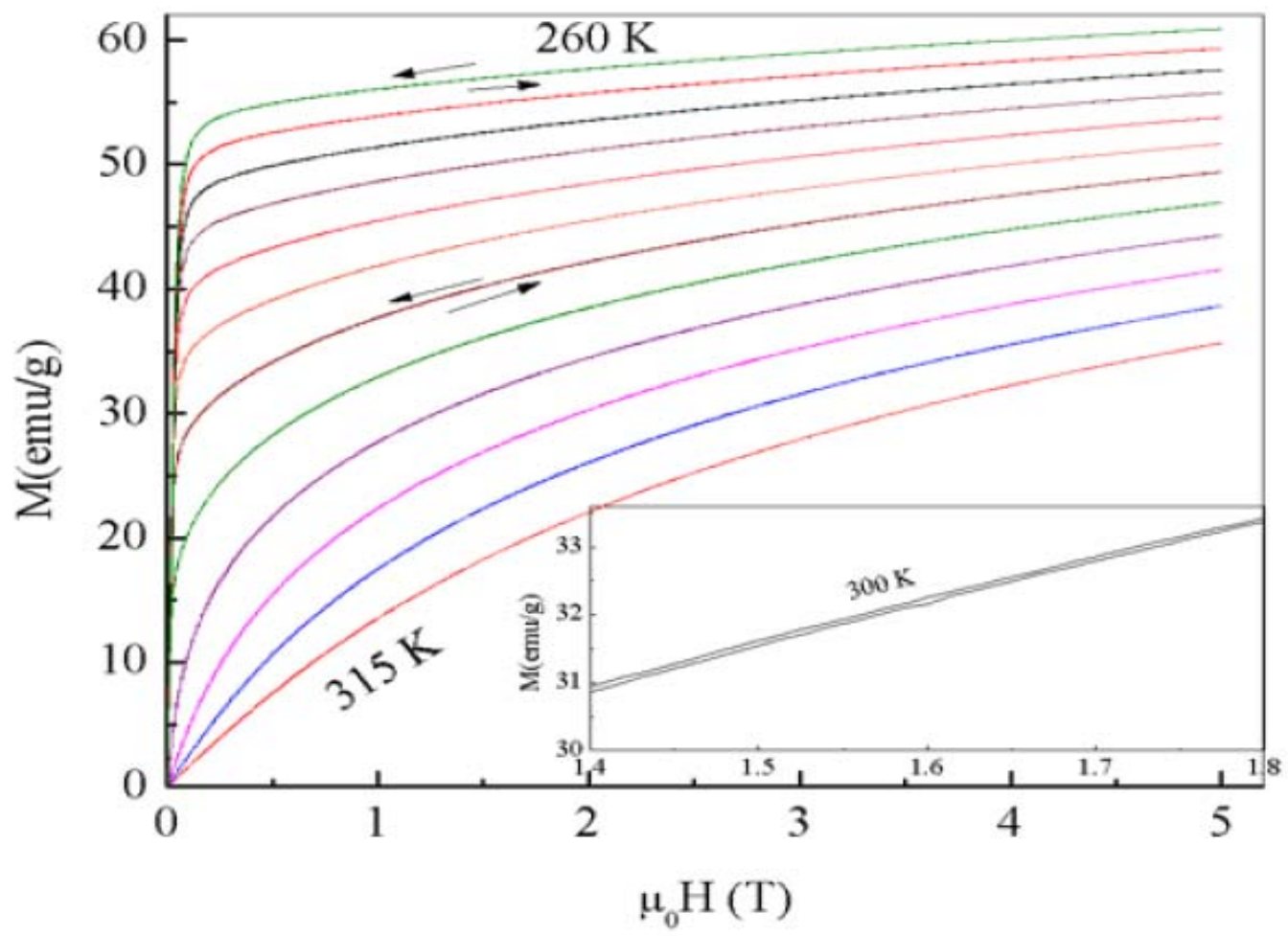

**Fig.10**: Magnetic field dependent magnetization of the $La_{0.7}Sr_{0.3}Mn_{0.93}Fe_{0.07}O_3$ in both ascending and descending field at 5 K intervals. The inset show enlarged view of 300 K magnetic isotherm between 1.4-1.8 T fields.

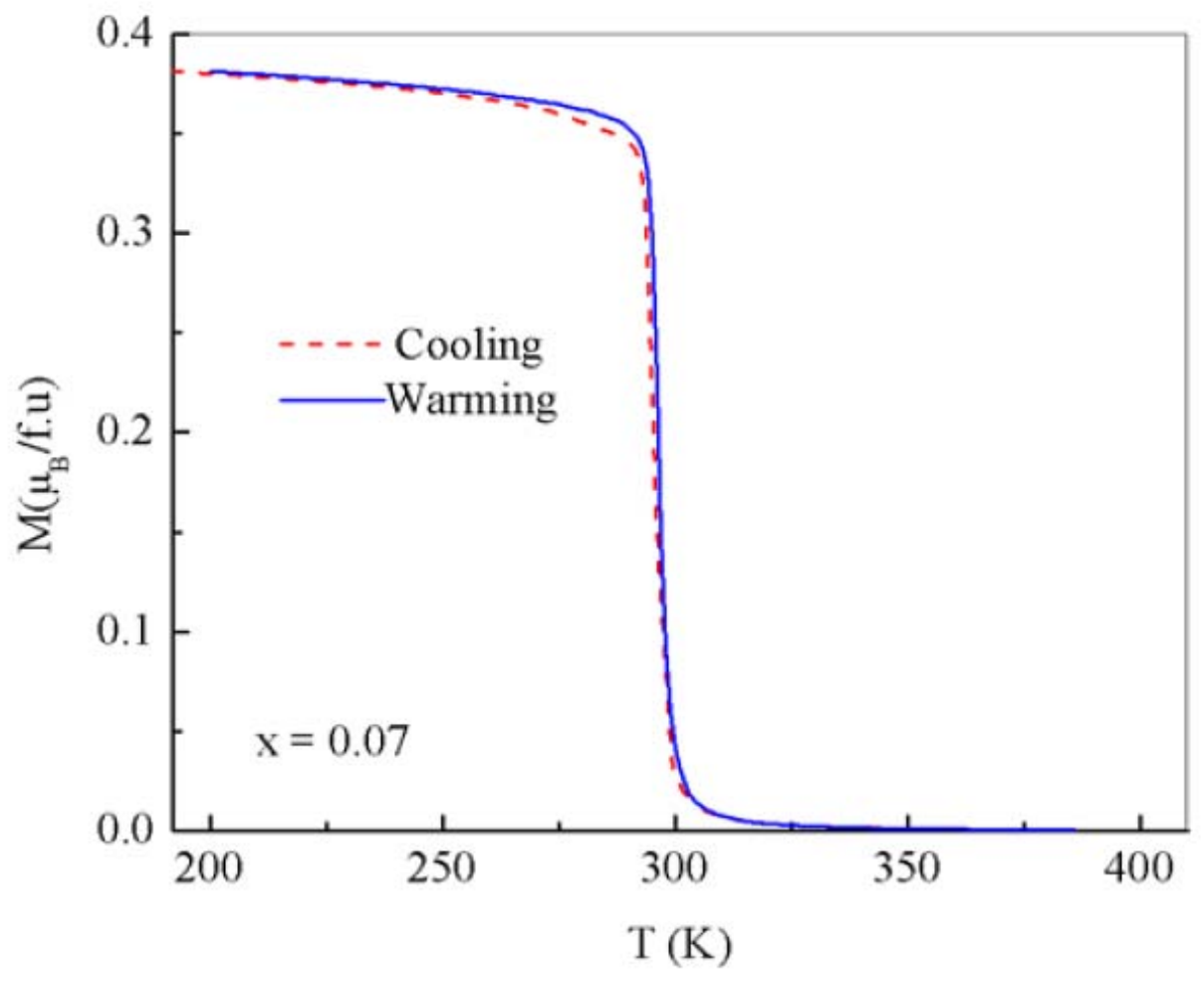

**Fig.11**: Temperature dependent magnetization of the $La_{0.7}Sr_{0.3}Mn_{0.93}Fe_{0.07}O_3$ in both cooling and warming cycle, around its $T_C$.

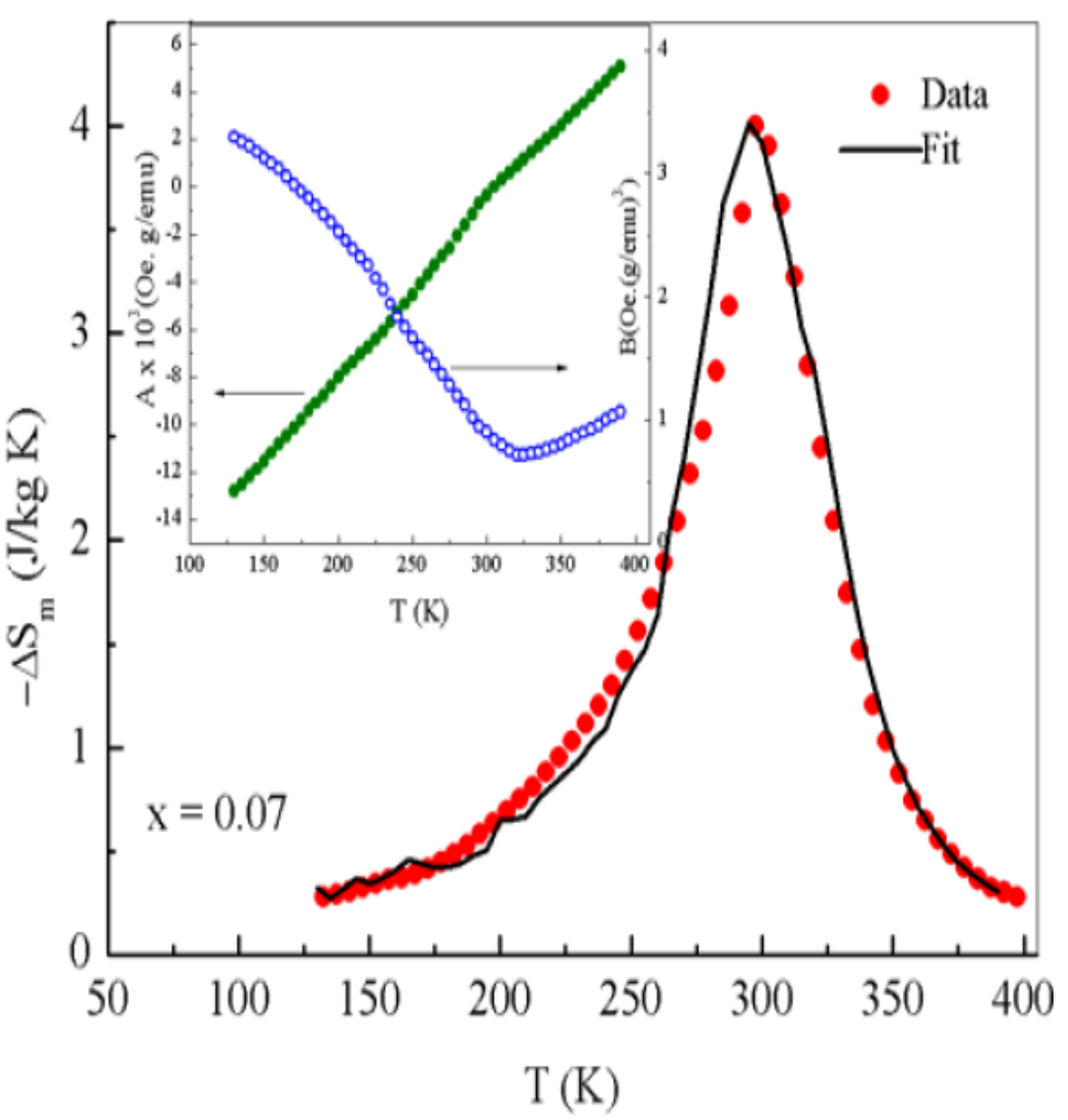

**Fig.12**: Temperature dependent $\Delta S_m$ of the $La_{0.7}Sr_{0.3}Mn_{0.93}Fe_{0.07}O_3$ sample along with theoretical curve (solid line). Inset: The temperature dependence A (left) & B (right ordinate) parameters of $H = AM+BM^3$.

## Table Captions:

| x | $T_C$ (K) | $\theta_p$ (K) | $p_{eff}$ (cal.) | $p_{eff}$ (expt.) | $M_s$ ($\mu_B$/f.u) (cal.) | $M_s$ ($\mu_B$/f.u) (expt.) |
|---|---|---|---|---|---|---|
| 0.05 | 343 | 347 | 4.64 | 6.3 | 3.75 | 3.52 |
| 0.07 | 296 | 307 | 4.66 | 4.9 | 3.77 | 3.35 |
| 0.10 | 260 | 283 | 4.70 | 5.4 | 3.80 | 3.20 |
| 0.15 | 180 | 217 | 4.74 | 6.2 | 3.85 | 2.92 |
| 0.20 | 94 | 172 | 4.80 | 7.1 | 3.90 | 1.43 |

**Table 1:** The determined $T_{C,}$ paramagnetic Curie temperature ($\theta_p$), paramagnetic effective magnetic moment ($P_{eff}$), and saturation magnetization ($M_s$) values of $La_{0.7}Sr_{0.3}Mn_{1-x}Fe_xO_3$ system. The abbreviations cal. and expt. represent calculated and experimental, respectively.

| Sample | $T_C$ (K) | $-\Delta S_m$ ($Jkg^{-1}K^{-1}$) ($\Delta H$= 2 T) | $-\Delta S_m$ ($Jkg^{-1}K^{-1}$) ($\Delta H$= 5 T) | Ref. |
|---|---|---|---|---|
| $La_{0.7}Sr_{0.3}MnO_3$ | 365 | 2.2 | 4.6 | 11 |
| $La_{0.7}Sr_{0.3}MnO_3$ | 374 | 1.8 | 2.5 (6T) | 28 |
| $La_{0.7}Sr_{0.3}MnO_3$ | 369 | 1.3 | 2.8 (6 T) | 25 |
| $La_{0.7}Sr_{0.3}Mn_{0.95}Fe_{0.05}O_3$ | 343 | 2.4 | 4.4 | Present |
| $La_{0.7}Sr_{0.3}Mn_{0.95}Al_{0.05}O_3$ | 332 | 1.4 | 3.4 | 11 |
| $La_{0.7}Sr_{0.3}Mn_{0.95}Ti_{0.05}O_3$ | 308 | 2.2 | 4.4 | 11 |
| $La_{0.7}Sr_{0.3}Mn_{0.95}Cu_{0.05}O_3$ | 350 | 1.9 (1.3T) | -- | 27 |
| $La_{0.7}Sr_{0.3}Mn_{0.93}Fe_{0.07}O_3$ | 296 | 2.2 | 4.0 | Present |
| $La_{0.7}Sr_{0.3}Mn_{0.90}Fe_{0.10}O_3$ | 260 | 1.7 | 3.1 | Present |
| $La_{0.7}Sr_{0.3}Mn_{0.90}Al_{0.10}O_3$ | 310 | 1.1 | 2.6 | 11 |
| $La_{0.7}Sr_{0.3}Mn_{0.95}Cu_{0.10}O_3$ | 350 | 2.0 (1.3T) | -- | 27 |
| $La_{0.7}Sr_{0.3}Mn_{0.85}Fe_{0.15}O_3$ | 180 | 1.4 | 2.6 | Present |
| $La_{0.7}Sr_{0.3}Mn_{0.80}Fe_{0.20}O_3$ | 94 | 0.6 | 1.3 | Present |
| $La_{0.7}Sr_{0.3}Mn_{0.80}Cr_{0.20}O_3$ | 286 | 1.2 | 2.6 (6 T) | 25 |
| $La_{0.7}Sr_{0.3}Mn_{0.60}Cr_{0.40}O_3$ | 242 | 0.5 | 1.2 (6 T) | 25 |

**Table 2:** The list of $\Delta S_m$ values under 2 T and 5 T for $La_{0.7}Sr_{0.3}Mn_{1-x}Fe_xO_3$ (x= 0.05, 0.07, 0.10, 0.15, & 0.20) and the related compounds along with their $T_C$. The magnetic fields different from the above values are given in brackets. Two hyphens indicate data not available. The references are listed in the last column.

| Sample | $T_h$-$T_c$ (K) ($\Delta$H= 2 T) | $T_h$-$T_c$ (K) ($\Delta$H= 5 T) | RCP (J/kg) ($\Delta$H= 2 T) | RCP(J/kg) ($\Delta$H= 5 T) | Ref. |
|---|---|---|---|---|---|
| $La_{0.7}Sr_{0.3}MnO_3$ | 30 | -- | 80 | -- | 11 |
| $La_{0.7}Sr_{0.3}MnO_3$ | 41 | -- | 69 | -- | 28 |
| $La_{0.7}Sr_{0.3}MnO_3$ | 29 | -- | 29 | -- | 25 |
| $La_{0.7}Sr_{0.3}Mn_{0.95}Fe_{0.05}O_3$ | 37 | 52 | 83 | 215 | Present |
| $La_{0.7}Sr_{0.3}Mn_{0.95}Al_{0.05}O_3$ | 58 | -- | 100 | -- | 11 |
| $La_{0.7}Sr_{0.3}Mn_{0.95}Ti_{0.05}O_3$ | 37 | -- | 90 | -- | 11 |
| $La_{0.7}Sr_{0.3}Mn_{0.93}Fe_{0.07}O_3$ | 41 | 59 | 88 | 225 | Present |
| $La_{0.7}Sr_{0.3}Mn_{0.90}Al_{0.10}O_3$ | 92 | -- | 109 | -- | 12 |
| $La_{0.7}Sr_{0.3}Mn_{0.90}Fe_{0.10}O_3$ | 50 | 68 | 83 | 192 | Present |
| $La_{0.7}Sr_{0.3}Mn_{0.85}Fe_{0.15}O_3$ | 74 | 103 | 97 | 255 | Present |
| $La_{0.7}Sr_{0.3}Mn_{0.80}Fe_{0.20}O_3$ | 98 | 130 | 90 | 168 | Present |
| $La_{0.7}Sr_{0.3}Mn_{0.80}Cr_{0.20}O_3$ | 49 | -- | 59 | -- | 25 |
| $La_{0.7}Sr_{0.3}Mn_{0.60}Cr_{0.40}O_3$ | 232 | -- | 110 | -- | 25 |

**Table 3:** The list of RCP and operating temperature range ($T_{hot}$-$T_{cold}$) values under 2 T and 5 T magnetic field for the $La_{0.7}Sr_{0.3}Mn_{1-x}Fe_xO_3$ (x= 0.05, 0.07, 0.10, 0.15, & 0.20) and the related compounds. Two hyphens indicate data not available. The references are listed in the last column.